\pdfoutput=1
\documentclass[reprint,aip,apl, twocolumn,floatfix,superscriptaddress]{revtex4-1}
\usepackage{graphicx}
\usepackage{dcolumn}
\usepackage{bm}
\usepackage{amsmath}
\usepackage{lineno}
\usepackage[english]{babel}
\usepackage[utf8]{inputenc}
\usepackage{color}
\usepackage{siunitx}
\usepackage{amsmath}
\usepackage{textcomp, gensymb}
\usepackage{upgreek}
\usepackage{hyperref}
\usepackage[T1]{fontenc}
\hypersetup{
    colorlinks=true,
    citecolor=magenta,
    linkcolor=blue,
    filecolor=magenta,      
    urlcolor=magenta,
    }

\DeclareUnicodeCharacter{2060}{\nolinebreak}
\begin{document}

\title{Reducing disorder in Ge quantum wells by using thick SiGe barriers}

\author{Davide Costa}
\affiliation{QuTech and Kavli Institute of Nanoscience, Delft University of Technology, Lorentzweg 1, 2628 CJ Delft, Netherlands}
\author{Lucas E. A. Stehouwer}
\affiliation{QuTech and Kavli Institute of Nanoscience, Delft University of Technology, Lorentzweg 1, 2628 CJ Delft, Netherlands}
\author{Yi Huang}
\affiliation{Condensed Matter Theory Center and Joint Quantum Institute, Department of Physics, University of Maryland, College Park, Maryland 20742, USA}
\author{Sara Martí-Sánchez}
\affiliation{Catalan Institute of Nanoscience and Nanotechnology (ICN2), CSIC and BIST, Campus UAB, Bellaterra, 08193 Barcelona, Catalonia, Spain}
\author{Davide Degli Esposti}
\affiliation{QuTech and Kavli Institute of Nanoscience, Delft University of Technology, Lorentzweg 1, 2628 CJ Delft, Netherlands}
\author{Jordi Arbiol}
\affiliation{Catalan Institute of Nanoscience and Nanotechnology (ICN2), CSIC and BIST, Campus UAB, Bellaterra, 08193 Barcelona, Catalonia, Spain}
\affiliation{ICREA, Pg. Lluís Companys 23, 08010 Barcelona, Catalonia, Spain}
\author{Giordano Scappucci}
\email{g.scappucci@tudelft.nl}
\affiliation{QuTech and Kavli Institute of Nanoscience, Delft University of Technology, Lorentzweg 1, 2628 CJ Delft, Netherlands}

\date{\today}

\begin{abstract}
We investigate the disorder properties of two-dimensional hole gases in Ge/SiGe heterostructures grown on Ge wafers, using thick SiGe barriers to mitigate the influence of the semiconductor-dielectric interface. Across several heterostructure field effect transistors we measure an average maximum mobility of $(4.4 \pm 0.2) \times 10^{6}~\mathrm{cm^2/Vs}$ at a saturation density of $(1.72 \pm 0.03) \times 10^{11}~\mathrm{cm^{-2}}$, corresponding to a long mean free path of $(30 \pm 1)~\mathrm{\upmu m}$. The highest measured mobility is $4.68 \times 10^{6}~\mathrm{cm^2/Vs}$. We identify uniform background impurities and interface roughness as the dominant scattering mechanisms limiting mobility in a representative device, and we evaluate a percolation-induced critical density of $(4.5 \pm 0.1)\times 10^{9} ~\mathrm{cm^{-2}}$. This low-disorder heterostructure, according to simulations, may support the electrostatic confinement of holes in gate-defined quantum dots. 
\end{abstract}

\maketitle

Holes confined in strained Ge quantum wells are a promising platform for quantum computation with gate-defined quantum dots \cite{scappucci2021germanium}. Beneficially, the low effective mass \cite{lodari2019light,terrazos2021theory} allows the definition of quantum dots using large electrodes \cite{lawrie2020quantum} and the sizeable spin-orbit coupling enables local qubit control that is fully electrical and fast \cite{hendrickx2020fast}. Recent progress in size and functionality of planar Ge qubit devices include the shared control of a 16 semiconductor quantum dot crossbar array \cite{borsoi2024shared}, a four qubit germanium quantum processor \cite{hendrickx2021four}, singlet-triplet qubits at low magnetic fields \cite{jirovec2021singlet}, sweet-spot qubit operation \cite{hendrickx2024sweet}, and hopping-based universal quantum logic \cite{wang2024operating}. All these demonstrations used Si as a substrate for the growth of Ge/SiGe heterostructures.
Previous work\cite{stehouwer2023germanium} has shown how moving from Si to Ge wafers, however, enables better Ge-rich SiGe strain-relaxed buffers (SRBs) because of the smaller lattice mismatch between substrate and strained Ge quantum well. These SRBs on a Ge wafer have a low threading dislocation density of $(6 \pm 1) \times 10^{5} ~\mathrm{cm^{-2}}$⁠, nearly one order of magnitude improvement compared to control SRBs on Si wafers\cite{sammak2019shallow}. The associated reduction in short range scattering allows for an improvement of the electrical performance of the two-dimensional hole gas (2DHG) in the quantum well, reaching a mobility of $(3.4 \pm 0.1) \times 10^{6} ~\mathrm{cm^2/Vs}$\cite{stehouwer2023germanium}.
In this Letter, we further explore the potential of Ge/SiGe heterostructures on Ge wafers by growing a thicker SiGe barrier to mitigate disorder from the semiconductor-dielectric interface. By combining Hall transport measurements and scattering theory, we demonstrate an improved mobility and percolation density, setting a benchmark for group IV semiconductors, whilst giving insights into the remaining sources of disorder. Schrödinger–Poisson simulations of quantum confined energy levels indicate that this heterostructure is suitable for hosting gate-defined hole quantum dots.

\begin{figure}
    \centering
	\includegraphics[width=85mm]{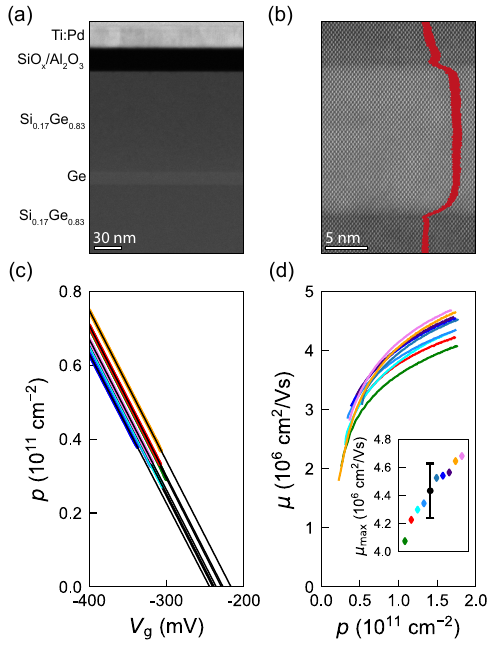}
	\caption{
    (a) HAADF-STEM image of the active layers of the Ge/SiGe heterostructure 
    field effect transistor. The 16~nm strained Ge quantum well is grown coherently on a Si$_{0.17}$Ge$_{0.83}$ strain-relaxed buffer. A 135~nm Si$_{0.17}$Ge$_{0.83}$ barrier separates the quantum well from  the dielectric stack. The dielectric stack comprises a thin native silicon oxide layer followed by a 30~nm Al$_{2}$O$_{3}$ film obtained by atomic layer deposition at $300 ~^\circ\mathrm{C}$. (b) Close-up image of the Ge quantum well with a superimposed average of the intensity profile (red curve). 
    (c) Hall density $p$ as a function of gate voltage $V_{\mathrm{g}}$ for 9 heterostructure field effect transistors from the same wafer (colored lines) and corresponding linear fits (black lines). (d) Hole mobility $\mu$ as a function of Hall density $p$ for the same heterostructure field effect transistors (colored lines) and distribution of maximum mobility $\mu_{\mathrm{max}}$ (inset). Maximum mobility from all the devices (diamonds), average value, and standard deviation (black) are shown.}
\label{fig:one}
\end{figure}

We grow a Ge/SiGe heterostructure on a Ge wafer by reduced-pressure chemical vapor deposition, following the same growth protocol described in Ref.~\cite{stehouwer2023germanium}. The epitaxial stack comprises a Si$_{0.17}$Ge$_{0.83}$ SRB by step grading, a compressively strained Ge quantum well, and a Si$_{0.17}$Ge$_{0.83}$ barrier passivated by a thin Si cap. Here, we increase the Si$_{0.17}$Ge$_{0.83}$ barrier thickness from $55~\mathrm{nm}$ \cite{stehouwer2023germanium} to $135~\mathrm{nm}$. We fabricate Hall-bar shaped heterostructure field effect transistors (H-FETs) for magnetotransport characterization by four-probes low-frequency lock-in techniques, as described in Ref.~\cite{sammak2019shallow}. Figure~\ref{fig:one}(a) shows a high angle annular dark field (HAADF) scanning transmission electron microscopy (STEM) image of the active layers of the H-FET, with no visible defects or dislocation. From the image in Fig.~\ref{fig:one}(b) we estimate a quantum well width $w = 16~\mathrm{nm}$ and a characteristic length-scale $4\tau$ of both the top and bottom interface of $2~\mathrm{nm}$, by fitting the intensity profile to a sigmoid model \cite{paquelet2022atomic, degli2024low} (see the supplementary material). However, the applicability of this model may be limited since we observe a slight accumulation of Si (or depletion of Ge) at both top and bottom interfaces of the quantum well, manifesting as darker lines in the HAADF-STEM image and therefore dips in the contrast profile.

We characterize disorder in this heterostructure by magnetotransport measurements across nine H-FETs in a dilution refrigerator equipped with a cryo-multiplexer \cite{paquelet2020multiplexed} (temperature $T=100 ~\mathrm{mK}$ measured at the mixing chamber). Applying a negative gate voltage $V_{\mathrm{g}}$ capacitively induces a 2DHG and controls the carrier density $p$ in the quantum well as shown in  Fig.~\ref{fig:one}(c). From the observed linear $p$-$V_{\mathrm{g}}$ relationships we obtain a capacitance per unit area $C$ of $(65.5 \pm 0.3) ~\mathrm{nF/cm^2}$  averaged across the H-FETs. The very small standard deviation (less than $0.5$\%) indicates highly uniform thickness and dielectric properties of the SiGe barrier and the SiO$_2$/Al$_2$O$_3$ layer on top. Furthermore, we extract the gate voltages $V_0$ for which the density extrapolates to zero [Fig.~\ref{fig:one}(c), black solid lines] and obtain an average value of $(-233.8 \pm 8.4) ~\mathrm{mV}$. Since $C$ is uniform throughout the devices, the tight distribution of $V_0$, with a standard deviation of $3.6$\%, is an indication of the very low disorder of the electrostatic potential landscape in the channel. Figure~\ref{fig:one}(d) shows the density-dependent mobility $\mu$. Whilst all devices achieve a similar saturation density $p_{\mathrm{sat}}$ of $(1.72 \pm 0.03) \times 10^{11} ~\mathrm{cm^{-2}}$, which is consistent with the uniform $C$, we do observe differences in the lowest measurable Hall density, which is $2.27 \times 10^{10} ~\mathrm{cm^{-2}}$ for the orange curve. These differences might arise from different contact resistances at low density across the devices. As a general trend, the mobility increases over the whole range of investigated density for all H-FETs. We observe a maximum mobility of $(4.4 \pm 0.2) \times 10^{6} ~\mathrm{cm^2/Vs}$ averaged across the 9 H-FETs [Fig.~\ref{fig:one}(d), inset] and measured at $p_{\mathrm{sat}}$, corresponding to an average mean free path of $(30 \pm 1) ~\mathrm{\upmu m}$. The largest mobility measured within the distribution is $4.68 \times 10^{6} ~\mathrm{cm^2/Vs}$, which represents a 9\% improvement over the previously highest reported hole mobility in Ge/SiGe heterostructures\cite{myronov2023holes}, setting a benchmark for group IV semiconductors. While the heterostructures in Ref~\cite{myronov2023holes} had a dislocation density twice as high, in the $10^6~\mathrm{cm^{-2}}$ range,  from being grown on a Si wafer, they exhibited less Si-Ge interdiffusion in the active region possibly due to a lower growth temperature (below $500 ~^{\circ}\mathrm{C}$) compared to our heterostructures.

\begin{figure}[t]
    \centering
	\includegraphics[width=80mm]{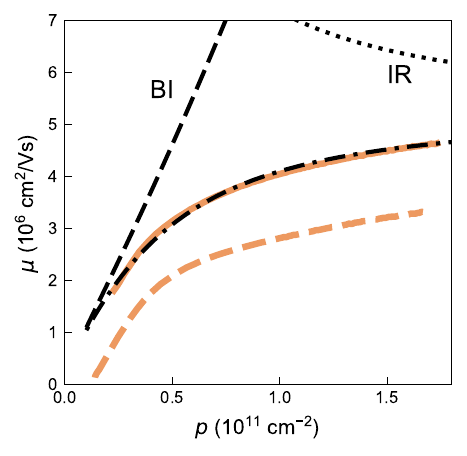}
	\caption{Mobility $\mu$ as a function of density $p$ from a representative heterostructure field effect transistor with a 135 nm thick SiGe barrier. The dashed-dotted black curve is a theoretical fit to the data considering scattering from a uniform background of charged impurities (BI) and interface roughness (IR). The individual contributions from BI and IR are shown as dashed black curve and dotted black curve, respectively. The dashed orange curve is from a similar heterostructure with a thinner SiGe barrier of 55~nm \cite{stehouwer2023germanium}.}
\label{fig:two}
\end{figure}

To gain insights into the scattering mechanisms in these high-mobility strained Ge quantum wells, we analyze in Fig.~\ref{fig:two} the mobility-density curve from the H-FET with the largest measured density range [orange curve, also plotted in Fig.~\ref{fig:one}(d)]. We use Boltzmann transport theory at $T=0 ~\mathrm{K}$ and calculate the mobility of the 2DHG in the Born approximation, taking into account the wavefunction of a finite potential well, the gate screening effect, and the local field correction following Ref.~\cite{huang2024understanding} (see the supplementary material). The fit to the theory (dashed-dotted black curve) shows good agreement with the experimental data and was calculated considering the contribution to the scattering rate from both uniform background charged impurities (BI) and interface roughness (IR). These individual contributions are shown as dashed black lines (BI) and dotted black lines (IR) in Fig.~\ref{fig:two}, and were obtained using a background impurity concentration of $N_{\mathrm{b}} = 3.8\times10^{13} ~\mathrm{cm^{-3}}$ and a surface roughness characterized by a typical height of $\Delta = 10$ {\AA} and lateral size of $\Lambda = 28$ {\AA}. To investigate how relevant is the disorder induced by the semiconductor-dielectric interface, we further evaluate the scattering rate contributed by remote charged impurities (RI). The mobility limited by RI has a quasi-cubic dependence (see the supplementary material) with the distance between the semiconductor-dielectric interface and the 2DHG in the quantum well, which is approximated by the SiGe barrier thickness. This suggests that if RI were the dominant scattering mechanism, increasing the barrier thickness from $55 ~\mathrm{nm}$ (dashed orange line, from Ref.~\cite{stehouwer2023germanium}) to $135 ~\mathrm{nm}$ (solid orange line) would result in more than one order of magnitude increase in mobility. However, the mobility increased by a factor of $\approx1.6$, from which we conclude that the dominant scattering mechanisms for this H-FET with a thick SiGe barrier are BI at low density and IR at high density, respectively.

\begin{figure}
  \centering
	\includegraphics[width=85mm]{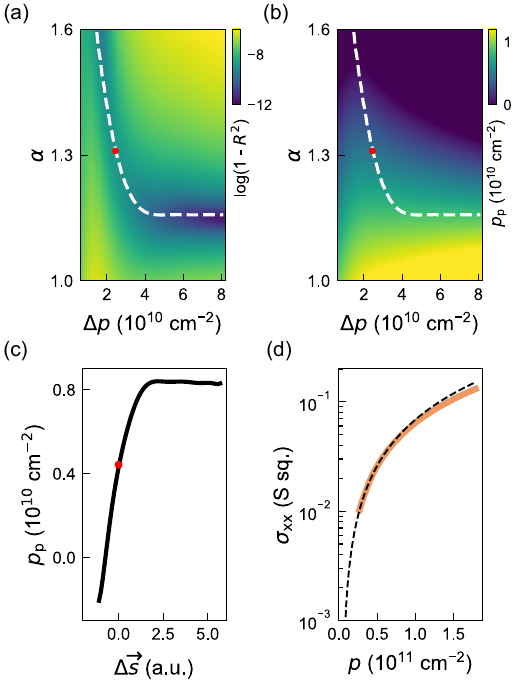}
	\caption{(a) Map of $\log(1 - R^2)$ as a function of variables $\alpha$ and $\Delta p$, where $R^2$ is the coefficient of determination obtained by fitting the density-dependent conductivity $\sigma_{\mathrm{xx}}(p)$ from the device in Fig.~\ref{fig:two} to the theoretical model $A(p - p_{\mathrm{p}})^{\alpha}$. Both $A$ and the percolation density $p_{\mathrm{p}}$ are free fitting parameters, while $\Delta p$ is the density range over which the data is fitted, starting from the minimum measured density. The dashed white curve $\overrightarrow{s}(\alpha,\Delta p)$ follows the local minima of $\log(1 - R^2)$. For $\alpha=1.31$, which is the critical exponent expected from percolation theory in 2D, the local minimum is found at $\Delta p = 2\times10^{10} ~\mathrm{cm^{-2}}$ (red dot). (b) Corresponding map of the fitted $p_{\mathrm{p}}$ as a function of $\alpha$ and $\Delta p$, with superimposed dashed white curve $\overrightarrow{s}(\alpha,\Delta p)$ and red dot as in (a). (c) Values of $p_{\mathrm{p}}$ as a function of the displacement $\Delta \overrightarrow{s}$ along the curve $\overrightarrow{s}(\alpha,\Delta p)$, relative to the position of the red dot. (d) Experimental $\sigma_{\mathrm{xx}}(p)$ curve (orange) and best fit to percolation theory in 2D (dashed black line). $\alpha$ and $\Delta p$ are fixed to the  values identifying the red dot in (a).}
\label{fig:three}
\end{figure}

We further investigate disorder in the low-density regime by analyzing, for the same H-FET, the density-dependent conductivity $\sigma_{xx}$ to determine the percolation density $p_{\mathrm{p}}$. The common method of extracting $p_{\mathrm{p}}$ is to fit the longitudinal conductivity $\sigma_{\mathrm{xx}}$ to the relationship $A(p - p_{\mathrm{p}})^{\alpha}$ \cite{last1971percolation,shklovskii1975percolation}, with coefficient $A$ and $p_{\mathrm{p}}$ as fitting parameters \cite{tracy2009observation} and exponent $\alpha$ fixed to the theoretical value of 1.31 from percolation theory in 2D \cite{fogelholm1980conductivity}. Two aspects of fitting percolation density, however, require attention: firstly, there are studies that consider the exponent $\alpha$ as a free fitting parameter \cite{kim2017annealing, myronov2023holes}. Secondly, since the theory is only valid in the low-density regime around $p_{\mathrm{p}}$, a meaningful estimate of $p_{\mathrm{p}}$ should account for its dependence on the density range $\Delta p$ over which the fit is performed \cite{lodari2021low, sabbagh2019quantum}. We systematically address these aspects by performing the following series of percolation fits: for a given value of $\alpha$, we fix the lower boundary of $\Delta p$ to the minimum measured Hall density of $2.27 \times 10^{10} ~\mathrm{cm^{-2}}$ and gradually increase the upper bound until it encompasses the full range under consideration. 

We assess the quality of the obtained percolation fits using the $R$-squared method and in Fig.~\ref{fig:three}(a) show $\log(1 - R^2)$  across the portion of the $\alpha - \Delta p$ plane investigated. The local minima are positioned along a continuous curve $\overrightarrow{s}(\alpha,\Delta p)$ (white dashed line), which identifies the region of best fit to the experimental data and crosses the theoretical 2D case of $\alpha = 1.31$ at a low $\Delta p = 2\times10^{10} ~\mathrm{cm^{-2}}$ (red circle). In Fig.~\ref{fig:three}(b), we show the obtained $p_{\mathrm{p}}$ values across the $\alpha - \Delta p$ plane and we use the curve $\overrightarrow{s}(\alpha,\Delta p)$ to refine the selection of $p_{\mathrm{p}}$ for a more reliable evaluation. The corresponding $p_{\mathrm{p}}$ values are shown in Fig.~\ref{fig:three}(c) as a function of displacement $\Delta \overrightarrow{s}$ along $\overrightarrow{s}(\alpha,\Delta p)$ and relative to the position of best fit for the theoretical 2D case (red circle). The fitted $p_{\mathrm{p}}$ shows an upper bound of about $8.5 \times 10^{9} ~\mathrm{cm^{-2}}$ for positive $\Delta \overrightarrow{s}$ at $ \alpha = 1.15$, whereas for negative displacements $\Delta \overrightarrow{s}$, i.e. $\alpha > 1.31$,  $p_{\mathrm{p}}$ may become arbitrarily small, and even take negative, and therefore unphysical, values for $\alpha > 1.5$.  
For the 2D theoretical case of a percolation-induced metal-insulator transition, characterized by a fixed universal exponent $\alpha = 1.31$, $p_{\mathrm{p}}$ assumes a value of $(4.5 \pm 0.1)\times 10^{9} ~\mathrm{cm^{-2}}$, with the experimental data and corresponding very good fit shown as a reference in Fig.~\ref{fig:three}(d).

\begin{figure}[t]
  \centering
	\includegraphics[width=85mm]{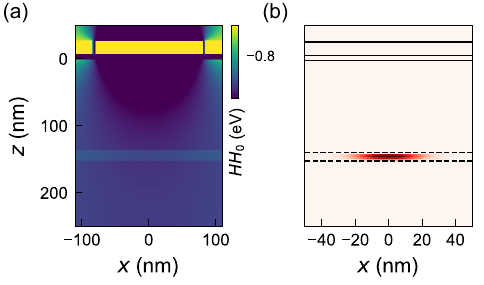}
	\caption{(a) Simulated ground state heavy hole (HH) band edge ($HH_{\mathrm{0}}$) as a function of the $x$ and $z$ coordinates. The strained Ge quantum well is positioned 135 $~\mathrm{nm}$ below the gate stack ($z < 0$). The plunger gate and the barrier gates are $160$ nm and $30$ nm wide, respectively. All gates are $30$ nm thick. (b) Hole population of the HH ground state, showing a quantum dot-like confinement.}
\label{fig:four}
\end{figure}

Having demonstrated the merits of a thick SiGe barrier for obtaining a very low-disordered 2DHG, we now assess the practicality of this design choice for confining holes in gate-defined quantum dots. We perform 2D Schrödinger–Poisson simulations using the \textsc{Nextnano} software package and show in Fig.~\ref{fig:four}(a) the edge of the first heavy-hole band ($HH_{\mathrm{0}}$). The top interface of the quantum well is positioned at $z=135 ~\mathrm{nm}$ and the gate stack (a plunger and two barrier gates with dielectric in between) is visible for $z<0 ~\mathrm{nm}$. Following calculations of the confined energy levels for the heavy-hole and light-hole states, we show in Fig.~\ref{fig:four}(b) that under appropriate voltages applied to the gates ($-0.58 ~\mathrm{V}$ and $-0.4 ~\mathrm{V}$ for plunger and barriers, respectively), only the first heavy-hole state is occupied in the quantum well. This supports the feasibility of fabricating quantum dots on this heterostructure.

In summary, we investigate Ge/SiGe heterostructures on Ge wafers and demonstrate an improved disorder in the Ge quantum well by using a thick SiGe barrier. The electrical performance of the 2DHG is limited by uniform background impurities and interface roughness and shows very high mobility and low percolation density, which we assessed with a critical evaluation of the fitting procedure. Additional insights into the structural and electrical properties of the heterostructure may be gained by using electron tomography to reconstruct the quantum well interface roughness and alloy disorder in three dimensions \cite{paysen2024three-dimensional}, and by evaluating the single-particle lifetime and quantum mobility \cite{dassarma2014mobility} from magnetotransport measurements. 
The high mobility reported recently in Ge/SiGe heterostructures with similar SiGe barrier thickness and alloy composition\cite{myronov2023holes}, despite their higher dislocation density because grown on a Si wafer, is very encouraging. These results suggest that much further improvement in the Ge/SiGe heterostructure on Ge wafers could be also achieved by purifying both Si and Ge gas precursors to suppress background contamination and by eventually lowering the growth temperature of the quantum well and barriers below $500~^{\circ}\mathrm{C}$ to suppress the Si and Ge interdiffusion at the quantum well interfaces.
We predict that quantum confinement of holes in gate-defined quantum dots is feasible in this heterostructure and we estimate that
dots of about $ 1/ \sqrt{p_\mathrm{p}} \approx 150\;\mathrm{nm}$ in size, informative about the average distance between impurities or other traps, could be essentially disorder-free. On the path to better performing spin-qubits, further measurements of charge noise in quantum dots will give insights into the dynamics of charge fluctuations that are not probed by magnetotransport experiments.

\vspace{\baselineskip}
See the supplementary material for fits of the intensity profile from the electron microscope images, notes on scattering theory for germanium quantum wells and conductivity-density curves for all measured H-FETs.

\vspace{\baselineskip}
We acknowledge Sankar Das Sarma for fruitful discussions. This work was supported by the Netherlands Organisation for Scientific Research (NWO/OCW), via the the Frontiers of Nanoscience program and a Projectruimte program. This research was supported by the European Union’s Horizon 2020 research and innovation programme under the grant agreement No. 951852 and No. 101069515. This research was sponsored in part by The Netherlands Ministry of Defence under Awards No. QuBits R23/009. The theoretical work for fitting of the mobility-density curves was supported by LPS, Laboratory of Physical Sciences. ICN2 acknowledges funding from Generalitat de Catalunya 2021SGR00457. ICN2 is supported by the Severo Ochoa program from Spanish MCIN/AEI (Grant No.: CEX2021-001214-S) and is funded by the CERCA Programme / Generalitat de Catalunya and ERDF funds from EU. We acknowledge support from CSIC Interdisciplinary Thematic Platform (PTI+) on Quantum Technologies (PTI-QTEP+). The views, conclusions, and recommendations contained in this document are those of the authors and are not necessarily endorsed nor should they be interpreted as representing the official policies, either expressed or implied, of The Netherlands Ministry of Defence. The Netherlands Ministry of Defence is authorized to reproduce and distribute reprints for Government purposes notwithstanding any copyright notation herein.

\section*{Author Declarations}
The authors have no conflict to disclose

\section*{Data availability}
The data sets supporting the findings of this study are openly available in 4TU Research Data at \url{https://doi.org/10.4121/a4fce2ed-262f-4db9-866b-661c0e002671}, Ref.~\cite{dataset}.


\begin{thebibliography}{27}%
\makeatletter
\providecommand \@ifxundefined [1]{%
 \@ifx{#1\undefined}
}%
\providecommand \@ifnum [1]{%
 \ifnum #1\expandafter \@firstoftwo
 \else \expandafter \@secondoftwo
 \fi
}%
\providecommand \@ifx [1]{%
 \ifx #1\expandafter \@firstoftwo
 \else \expandafter \@secondoftwo
 \fi
}%
\providecommand \natexlab [1]{#1}%
\providecommand \enquote  [1]{``#1''}%
\providecommand \bibnamefont  [1]{#1}%
\providecommand \bibfnamefont [1]{#1}%
\providecommand \citenamefont [1]{#1}%
\providecommand \href@noop [0]{\@secondoftwo}%
\providecommand \href [0]{\begingroup \@sanitize@url \@href}%
\providecommand \@href[1]{\@@startlink{#1}\@@href}%
\providecommand \@@href[1]{\endgroup#1\@@endlink}%
\providecommand \@sanitize@url [0]{\catcode `\\12\catcode `\$12\catcode `\&12\catcode `\#12\catcode `\^12\catcode `\_12\catcode `\%12\relax}%
\providecommand \@@startlink[1]{}%
\providecommand \@@endlink[0]{}%
\providecommand \url  [0]{\begingroup\@sanitize@url \@url }%
\providecommand \@url [1]{\endgroup\@href {#1}{\urlprefix }}%
\providecommand \urlprefix  [0]{URL }%
\providecommand \Eprint [0]{\href }%
\providecommand \doibase [0]{http://dx.doi.org/}%
\providecommand \selectlanguage [0]{\@gobble}%
\providecommand \bibinfo  [0]{\@secondoftwo}%
\providecommand \bibfield  [0]{\@secondoftwo}%
\providecommand \translation [1]{[#1]}%
\providecommand \BibitemOpen [0]{}%
\providecommand \bibitemStop [0]{}%
\providecommand \bibitemNoStop [0]{.\EOS\space}%
\providecommand \EOS [0]{\spacefactor3000\relax}%
\providecommand \BibitemShut  [1]{\csname bibitem#1\endcsname}%
\let\auto@bib@innerbib\@empty
\bibitem [{\citenamefont {Scappucci}\ \emph {et~al.}(2021)\citenamefont {Scappucci}, \citenamefont {Kloeffel}, \citenamefont {Zwanenburg}, \citenamefont {Loss}, \citenamefont {Myronov}, \citenamefont {Zhang}, \citenamefont {De~Franceschi}, \citenamefont {Katsaros},\ and\ \citenamefont {Veldhorst}}]{scappucci2021germanium}%
  \BibitemOpen
  \bibfield  {author} {\bibinfo {author} {\bibfnamefont {G.}~\bibnamefont {Scappucci}}, \bibinfo {author} {\bibfnamefont {C.}~\bibnamefont {Kloeffel}}, \bibinfo {author} {\bibfnamefont {F.~A.}\ \bibnamefont {Zwanenburg}}, \bibinfo {author} {\bibfnamefont {D.}~\bibnamefont {Loss}}, \bibinfo {author} {\bibfnamefont {M.}~\bibnamefont {Myronov}}, \bibinfo {author} {\bibfnamefont {J.-J.}\ \bibnamefont {Zhang}}, \bibinfo {author} {\bibfnamefont {S.}~\bibnamefont {De~Franceschi}}, \bibinfo {author} {\bibfnamefont {G.}~\bibnamefont {Katsaros}}, \ and\ \bibinfo {author} {\bibfnamefont {M.}~\bibnamefont {Veldhorst}},\ }\href {\doibase https://doi.org/10.1038/s41578-020-00262-z} {\bibfield  {journal} {\bibinfo  {journal} {Nature Reviews Materials}\ }\textbf {\bibinfo {volume} {6}},\ \bibinfo {pages} {926} (\bibinfo {year} {2021})}\BibitemShut {NoStop}%
\bibitem [{\citenamefont {Lodari}\ \emph {et~al.}(2019)\citenamefont {Lodari}, \citenamefont {Tosato}, \citenamefont {Sabbagh}, \citenamefont {Schubert}, \citenamefont {Capellini}, \citenamefont {Sammak}, \citenamefont {Veldhorst},\ and\ \citenamefont {Scappucci}}]{lodari2019light}%
  \BibitemOpen
  \bibfield  {author} {\bibinfo {author} {\bibfnamefont {M.}~\bibnamefont {Lodari}}, \bibinfo {author} {\bibfnamefont {A.}~\bibnamefont {Tosato}}, \bibinfo {author} {\bibfnamefont {D.}~\bibnamefont {Sabbagh}}, \bibinfo {author} {\bibfnamefont {M.~A.}\ \bibnamefont {Schubert}}, \bibinfo {author} {\bibfnamefont {G.}~\bibnamefont {Capellini}}, \bibinfo {author} {\bibfnamefont {A.}~\bibnamefont {Sammak}}, \bibinfo {author} {\bibfnamefont {M.}~\bibnamefont {Veldhorst}}, \ and\ \bibinfo {author} {\bibfnamefont {G.}~\bibnamefont {Scappucci}},\ }\href {\doibase 10.1103/PhysRevB.100.041304} {\bibfield  {journal} {\bibinfo  {journal} {Phys. Rev. B}\ }\textbf {\bibinfo {volume} {100}},\ \bibinfo {pages} {041304} (\bibinfo {year} {2019})}\BibitemShut {NoStop}%
\bibitem [{\citenamefont {Terrazos}\ \emph {et~al.}(2021)\citenamefont {Terrazos}, \citenamefont {Marcellina}, \citenamefont {Wang}, \citenamefont {Coppersmith}, \citenamefont {Friesen}, \citenamefont {Hamilton}, \citenamefont {Hu}, \citenamefont {Koiller}, \citenamefont {Saraiva}, \citenamefont {Culcer},\ and\ \citenamefont {Capaz}}]{terrazos2021theory}%
  \BibitemOpen
  \bibfield  {author} {\bibinfo {author} {\bibfnamefont {L.~A.}\ \bibnamefont {Terrazos}}, \bibinfo {author} {\bibfnamefont {E.}~\bibnamefont {Marcellina}}, \bibinfo {author} {\bibfnamefont {Z.}~\bibnamefont {Wang}}, \bibinfo {author} {\bibfnamefont {S.~N.}\ \bibnamefont {Coppersmith}}, \bibinfo {author} {\bibfnamefont {M.}~\bibnamefont {Friesen}}, \bibinfo {author} {\bibfnamefont {A.~R.}\ \bibnamefont {Hamilton}}, \bibinfo {author} {\bibfnamefont {X.}~\bibnamefont {Hu}}, \bibinfo {author} {\bibfnamefont {B.}~\bibnamefont {Koiller}}, \bibinfo {author} {\bibfnamefont {A.~L.}\ \bibnamefont {Saraiva}}, \bibinfo {author} {\bibfnamefont {D.}~\bibnamefont {Culcer}}, \ and\ \bibinfo {author} {\bibfnamefont {R.~B.}\ \bibnamefont {Capaz}},\ }\href {\doibase https://doi.org/10.1103/PhysRevB.103.125201} {\bibfield  {journal} {\bibinfo  {journal} {Physical Review B}\ }\textbf {\bibinfo {volume} {103}},\ \bibinfo {pages} {125201} (\bibinfo {year} {2021})}\BibitemShut {NoStop}%
\bibitem [{\citenamefont {Lawrie}\ \emph {et~al.}(2020)\citenamefont {Lawrie}, \citenamefont {Eenink}, \citenamefont {Hendrickx}, \citenamefont {Boter}, \citenamefont {Petit}, \citenamefont {Amitonov}, \citenamefont {Lodari}, \citenamefont {Paquelet~Wuetz}, \citenamefont {Volk}, \citenamefont {Philips} \emph {et~al.}}]{lawrie2020quantum}%
  \BibitemOpen
  \bibfield  {author} {\bibinfo {author} {\bibfnamefont {W.}~\bibnamefont {Lawrie}}, \bibinfo {author} {\bibfnamefont {H.}~\bibnamefont {Eenink}}, \bibinfo {author} {\bibfnamefont {N.}~\bibnamefont {Hendrickx}}, \bibinfo {author} {\bibfnamefont {J.}~\bibnamefont {Boter}}, \bibinfo {author} {\bibfnamefont {L.}~\bibnamefont {Petit}}, \bibinfo {author} {\bibfnamefont {S.}~\bibnamefont {Amitonov}}, \bibinfo {author} {\bibfnamefont {M.}~\bibnamefont {Lodari}}, \bibinfo {author} {\bibfnamefont {B.}~\bibnamefont {Paquelet~Wuetz}}, \bibinfo {author} {\bibfnamefont {C.}~\bibnamefont {Volk}}, \bibinfo {author} {\bibfnamefont {S.}~\bibnamefont {Philips}},  \emph {et~al.},\ }\href {\doibase https://doi.org/10.1063/5.0002013} {\bibfield  {journal} {\bibinfo  {journal} {Applied Physics Letters}\ }\textbf {\bibinfo {volume} {116}} (\bibinfo {year} {2020}),\ https://doi.org/10.1063/5.0002013}\BibitemShut {NoStop}%
\bibitem [{\citenamefont {Hendrickx}\ \emph {et~al.}(2020)\citenamefont {Hendrickx}, \citenamefont {Franke}, \citenamefont {Sammak}, \citenamefont {Scappucci},\ and\ \citenamefont {Veldhorst}}]{hendrickx2020fast}%
  \BibitemOpen
  \bibfield  {author} {\bibinfo {author} {\bibfnamefont {N.}~\bibnamefont {Hendrickx}}, \bibinfo {author} {\bibfnamefont {D.}~\bibnamefont {Franke}}, \bibinfo {author} {\bibfnamefont {A.}~\bibnamefont {Sammak}}, \bibinfo {author} {\bibfnamefont {G.}~\bibnamefont {Scappucci}}, \ and\ \bibinfo {author} {\bibfnamefont {M.}~\bibnamefont {Veldhorst}},\ }\href {\doibase https://doi.org/10.1038/s41586-019-1919-3} {\bibfield  {journal} {\bibinfo  {journal} {Nature}\ }\textbf {\bibinfo {volume} {577}},\ \bibinfo {pages} {487} (\bibinfo {year} {2020})}\BibitemShut {NoStop}%
\bibitem [{\citenamefont {Borsoi}\ \emph {et~al.}(2024)\citenamefont {Borsoi}, \citenamefont {Hendrickx}, \citenamefont {John}, \citenamefont {Meyer}, \citenamefont {Motz}, \citenamefont {van Riggelen}, \citenamefont {Sammak}, \citenamefont {de~Snoo}, \citenamefont {Scappucci},\ and\ \citenamefont {Veldhorst}}]{borsoi2024shared}%
  \BibitemOpen
  \bibfield  {author} {\bibinfo {author} {\bibfnamefont {F.}~\bibnamefont {Borsoi}}, \bibinfo {author} {\bibfnamefont {N.~W.}\ \bibnamefont {Hendrickx}}, \bibinfo {author} {\bibfnamefont {V.}~\bibnamefont {John}}, \bibinfo {author} {\bibfnamefont {M.}~\bibnamefont {Meyer}}, \bibinfo {author} {\bibfnamefont {S.}~\bibnamefont {Motz}}, \bibinfo {author} {\bibfnamefont {F.}~\bibnamefont {van Riggelen}}, \bibinfo {author} {\bibfnamefont {A.}~\bibnamefont {Sammak}}, \bibinfo {author} {\bibfnamefont {S.~L.}\ \bibnamefont {de~Snoo}}, \bibinfo {author} {\bibfnamefont {G.}~\bibnamefont {Scappucci}}, \ and\ \bibinfo {author} {\bibfnamefont {M.}~\bibnamefont {Veldhorst}},\ }\href {\doibase https://doi.org/10.1038/s41565-023-01491-3} {\bibfield  {journal} {\bibinfo  {journal} {Nature Nanotechnology}\ }\textbf {\bibinfo {volume} {19}},\ \bibinfo {pages} {21} (\bibinfo {year} {2024})}\BibitemShut {NoStop}%
\bibitem [{\citenamefont {Hendrickx}\ \emph {et~al.}(2021)\citenamefont {Hendrickx}, \citenamefont {Lawrie}, \citenamefont {Russ}, \citenamefont {van Riggelen}, \citenamefont {de~Snoo}, \citenamefont {Schouten}, \citenamefont {Sammak}, \citenamefont {Scappucci},\ and\ \citenamefont {Veldhorst}}]{hendrickx2021four}%
  \BibitemOpen
  \bibfield  {author} {\bibinfo {author} {\bibfnamefont {N.~W.}\ \bibnamefont {Hendrickx}}, \bibinfo {author} {\bibfnamefont {W.~I.}\ \bibnamefont {Lawrie}}, \bibinfo {author} {\bibfnamefont {M.}~\bibnamefont {Russ}}, \bibinfo {author} {\bibfnamefont {F.}~\bibnamefont {van Riggelen}}, \bibinfo {author} {\bibfnamefont {S.~L.}\ \bibnamefont {de~Snoo}}, \bibinfo {author} {\bibfnamefont {R.~N.}\ \bibnamefont {Schouten}}, \bibinfo {author} {\bibfnamefont {A.}~\bibnamefont {Sammak}}, \bibinfo {author} {\bibfnamefont {G.}~\bibnamefont {Scappucci}}, \ and\ \bibinfo {author} {\bibfnamefont {M.}~\bibnamefont {Veldhorst}},\ }\href {\doibase https://doi.org/10.1038/s41586-021-03332-6} {\bibfield  {journal} {\bibinfo  {journal} {Nature}\ }\textbf {\bibinfo {volume} {591}},\ \bibinfo {pages} {580} (\bibinfo {year} {2021})}\BibitemShut {NoStop}%
\bibitem [{\citenamefont {Jirovec}\ \emph {et~al.}(2021)\citenamefont {Jirovec}, \citenamefont {Hofmann}, \citenamefont {Ballabio}, \citenamefont {Mutter}, \citenamefont {Tavani}, \citenamefont {Botifoll}, \citenamefont {Crippa}, \citenamefont {Kukucka}, \citenamefont {Sagi}, \citenamefont {Martins} \emph {et~al.}}]{jirovec2021singlet}%
  \BibitemOpen
  \bibfield  {author} {\bibinfo {author} {\bibfnamefont {D.}~\bibnamefont {Jirovec}}, \bibinfo {author} {\bibfnamefont {A.}~\bibnamefont {Hofmann}}, \bibinfo {author} {\bibfnamefont {A.}~\bibnamefont {Ballabio}}, \bibinfo {author} {\bibfnamefont {P.~M.}\ \bibnamefont {Mutter}}, \bibinfo {author} {\bibfnamefont {G.}~\bibnamefont {Tavani}}, \bibinfo {author} {\bibfnamefont {M.}~\bibnamefont {Botifoll}}, \bibinfo {author} {\bibfnamefont {A.}~\bibnamefont {Crippa}}, \bibinfo {author} {\bibfnamefont {J.}~\bibnamefont {Kukucka}}, \bibinfo {author} {\bibfnamefont {O.}~\bibnamefont {Sagi}}, \bibinfo {author} {\bibfnamefont {F.}~\bibnamefont {Martins}},  \emph {et~al.},\ }\href {\doibase https://doi.org/10.1038/s41563-021-01022-2} {\bibfield  {journal} {\bibinfo  {journal} {Nature Materials}\ }\textbf {\bibinfo {volume} {20}},\ \bibinfo {pages} {1106} (\bibinfo {year} {2021})}\BibitemShut {NoStop}%
\bibitem [{\citenamefont {Hendrickx}\ \emph {et~al.}(2024)\citenamefont {Hendrickx}, \citenamefont {Massai}, \citenamefont {Mergenthaler}, \citenamefont {Schupp}, \citenamefont {Paredes}, \citenamefont {Bedell}, \citenamefont {Salis},\ and\ \citenamefont {Fuhrer}}]{hendrickx2024sweet}%
  \BibitemOpen
  \bibfield  {author} {\bibinfo {author} {\bibfnamefont {N.}~\bibnamefont {Hendrickx}}, \bibinfo {author} {\bibfnamefont {L.}~\bibnamefont {Massai}}, \bibinfo {author} {\bibfnamefont {M.}~\bibnamefont {Mergenthaler}}, \bibinfo {author} {\bibfnamefont {F.}~\bibnamefont {Schupp}}, \bibinfo {author} {\bibfnamefont {S.}~\bibnamefont {Paredes}}, \bibinfo {author} {\bibfnamefont {S.}~\bibnamefont {Bedell}}, \bibinfo {author} {\bibfnamefont {G.}~\bibnamefont {Salis}}, \ and\ \bibinfo {author} {\bibfnamefont {A.}~\bibnamefont {Fuhrer}},\ }\href {\doibase https://doi.org/10.1038/s41563-024-01857-5} {\bibfield  {journal} {\bibinfo  {journal} {Nature Materials}\ ,\ \bibinfo {pages} {1}} (\bibinfo {year} {2024})}\BibitemShut {NoStop}%
\bibitem [{\citenamefont {Wang}\ \emph {et~al.}(2024)\citenamefont {Wang}, \citenamefont {John}, \citenamefont {Tidjani}, \citenamefont {Yu}, \citenamefont {Ivlev}, \citenamefont {D{\'e}prez}, \citenamefont {van Riggelen-Doelman}, \citenamefont {Woods}, \citenamefont {Hendrickx}, \citenamefont {Lawrie} \emph {et~al.}}]{wang2024operating}%
  \BibitemOpen
  \bibfield  {author} {\bibinfo {author} {\bibfnamefont {C.-A.}\ \bibnamefont {Wang}}, \bibinfo {author} {\bibfnamefont {V.}~\bibnamefont {John}}, \bibinfo {author} {\bibfnamefont {H.}~\bibnamefont {Tidjani}}, \bibinfo {author} {\bibfnamefont {C.~X.}\ \bibnamefont {Yu}}, \bibinfo {author} {\bibfnamefont {A.~S.}\ \bibnamefont {Ivlev}}, \bibinfo {author} {\bibfnamefont {C.}~\bibnamefont {D{\'e}prez}}, \bibinfo {author} {\bibfnamefont {F.}~\bibnamefont {van Riggelen-Doelman}}, \bibinfo {author} {\bibfnamefont {B.~D.}\ \bibnamefont {Woods}}, \bibinfo {author} {\bibfnamefont {N.~W.}\ \bibnamefont {Hendrickx}}, \bibinfo {author} {\bibfnamefont {W.~I.}\ \bibnamefont {Lawrie}},  \emph {et~al.},\ }\href {\doibase https://doi.org/10.1126/science.ado5915} {\bibfield  {journal} {\bibinfo  {journal} {Science}\ }\textbf {\bibinfo {volume} {385}},\ \bibinfo {pages} {447} (\bibinfo {year} {2024})}\BibitemShut {NoStop}%
\bibitem [{\citenamefont {Stehouwer}\ \emph {et~al.}(2023)\citenamefont {Stehouwer}, \citenamefont {Tosato}, \citenamefont {Degli~Esposti}, \citenamefont {Costa}, \citenamefont {Veldhorst}, \citenamefont {Sammak},\ and\ \citenamefont {Scappucci}}]{stehouwer2023germanium}%
  \BibitemOpen
  \bibfield  {author} {\bibinfo {author} {\bibfnamefont {L.~E.}\ \bibnamefont {Stehouwer}}, \bibinfo {author} {\bibfnamefont {A.}~\bibnamefont {Tosato}}, \bibinfo {author} {\bibfnamefont {D.}~\bibnamefont {Degli~Esposti}}, \bibinfo {author} {\bibfnamefont {D.}~\bibnamefont {Costa}}, \bibinfo {author} {\bibfnamefont {M.}~\bibnamefont {Veldhorst}}, \bibinfo {author} {\bibfnamefont {A.}~\bibnamefont {Sammak}}, \ and\ \bibinfo {author} {\bibfnamefont {G.}~\bibnamefont {Scappucci}},\ }\href {\doibase https://doi.org/10.1063/5.0158262} {\bibfield  {journal} {\bibinfo  {journal} {Applied Physics Letters}\ }\textbf {\bibinfo {volume} {123}} (\bibinfo {year} {2023}),\ https://doi.org/10.1063/5.0158262}\BibitemShut {NoStop}%
\bibitem [{\citenamefont {Sammak}\ \emph {et~al.}(2019)\citenamefont {Sammak}, \citenamefont {Sabbagh}, \citenamefont {Hendrickx}, \citenamefont {Lodari}, \citenamefont {Paquelet~Wuetz}, \citenamefont {Tosato}, \citenamefont {Yeoh}, \citenamefont {Bollani}, \citenamefont {Virgilio}, \citenamefont {Schubert} \emph {et~al.}}]{sammak2019shallow}%
  \BibitemOpen
  \bibfield  {author} {\bibinfo {author} {\bibfnamefont {A.}~\bibnamefont {Sammak}}, \bibinfo {author} {\bibfnamefont {D.}~\bibnamefont {Sabbagh}}, \bibinfo {author} {\bibfnamefont {N.~W.}\ \bibnamefont {Hendrickx}}, \bibinfo {author} {\bibfnamefont {M.}~\bibnamefont {Lodari}}, \bibinfo {author} {\bibfnamefont {B.}~\bibnamefont {Paquelet~Wuetz}}, \bibinfo {author} {\bibfnamefont {A.}~\bibnamefont {Tosato}}, \bibinfo {author} {\bibfnamefont {L.}~\bibnamefont {Yeoh}}, \bibinfo {author} {\bibfnamefont {M.}~\bibnamefont {Bollani}}, \bibinfo {author} {\bibfnamefont {M.}~\bibnamefont {Virgilio}}, \bibinfo {author} {\bibfnamefont {M.~A.}\ \bibnamefont {Schubert}},  \emph {et~al.},\ }\href {\doibase https://doi.org/10.1002/adfm.201807613} {\bibfield  {journal} {\bibinfo  {journal} {Advanced Functional Materials}\ }\textbf {\bibinfo {volume} {29}},\ \bibinfo {pages} {1807613} (\bibinfo {year} {2019})}\BibitemShut {NoStop}%
\bibitem [{\citenamefont {Paquelet~Wuetz}\ \emph {et~al.}(2022)\citenamefont {Paquelet~Wuetz}, \citenamefont {Losert}, \citenamefont {Koelling}, \citenamefont {Stehouwer}, \citenamefont {Zwerver}, \citenamefont {Philips}, \citenamefont {Mądzik}, \citenamefont {Xue}, \citenamefont {Zheng}, \citenamefont {Lodari}, \citenamefont {Amitonov}, \citenamefont {Samkharadze}, \citenamefont {Sammak}, \citenamefont {Vandersypen}, \citenamefont {Rahman}, \citenamefont {Coppersmith}, \citenamefont {Moutanabbir}, \citenamefont {Friesen},\ and\ \citenamefont {Scappucci}}]{paquelet2022atomic}%
  \BibitemOpen
  \bibfield  {author} {\bibinfo {author} {\bibfnamefont {B.}~\bibnamefont {Paquelet~Wuetz}}, \bibinfo {author} {\bibfnamefont {M.~P.}\ \bibnamefont {Losert}}, \bibinfo {author} {\bibfnamefont {S.}~\bibnamefont {Koelling}}, \bibinfo {author} {\bibfnamefont {L.~E.~A.}\ \bibnamefont {Stehouwer}}, \bibinfo {author} {\bibfnamefont {A.-M.~J.}\ \bibnamefont {Zwerver}}, \bibinfo {author} {\bibfnamefont {S.~G.~J.}\ \bibnamefont {Philips}}, \bibinfo {author} {\bibfnamefont {M.~T.}\ \bibnamefont {Mądzik}}, \bibinfo {author} {\bibfnamefont {X.}~\bibnamefont {Xue}}, \bibinfo {author} {\bibfnamefont {G.}~\bibnamefont {Zheng}}, \bibinfo {author} {\bibfnamefont {M.}~\bibnamefont {Lodari}}, \bibinfo {author} {\bibfnamefont {S.~V.}\ \bibnamefont {Amitonov}}, \bibinfo {author} {\bibfnamefont {N.}~\bibnamefont {Samkharadze}}, \bibinfo {author} {\bibfnamefont {A.}~\bibnamefont {Sammak}}, \bibinfo {author} {\bibfnamefont {L.~M.~K.}\ \bibnamefont {Vandersypen}}, \bibinfo {author} {\bibfnamefont {R.}~\bibnamefont {Rahman}},
  \bibinfo {author} {\bibfnamefont {S.~N.}\ \bibnamefont {Coppersmith}}, \bibinfo {author} {\bibfnamefont {O.}~\bibnamefont {Moutanabbir}}, \bibinfo {author} {\bibfnamefont {M.}~\bibnamefont {Friesen}}, \ and\ \bibinfo {author} {\bibfnamefont {G.}~\bibnamefont {Scappucci}},\ }\href {\doibase 10.1038/s41467-022-35458-0} {\bibfield  {journal} {\bibinfo  {journal} {Nature Communications}\ }\textbf {\bibinfo {volume} {13}},\ \bibinfo {pages} {7730} (\bibinfo {year} {2022})}\BibitemShut {NoStop}%
\bibitem [{\citenamefont {Degli~Esposti}\ \emph {et~al.}(2024)\citenamefont {Degli~Esposti}, \citenamefont {Stehouwer}, \citenamefont {G{\"u}l}, \citenamefont {Samkharadze}, \citenamefont {D{\'e}prez}, \citenamefont {Meyer}, \citenamefont {Meijer}, \citenamefont {Tryputen}, \citenamefont {Karwal}, \citenamefont {Botifoll} \emph {et~al.}}]{degli2024low}%
  \BibitemOpen
  \bibfield  {author} {\bibinfo {author} {\bibfnamefont {D.}~\bibnamefont {Degli~Esposti}}, \bibinfo {author} {\bibfnamefont {L.~E.}\ \bibnamefont {Stehouwer}}, \bibinfo {author} {\bibfnamefont {{\"O}.}~\bibnamefont {G{\"u}l}}, \bibinfo {author} {\bibfnamefont {N.}~\bibnamefont {Samkharadze}}, \bibinfo {author} {\bibfnamefont {C.}~\bibnamefont {D{\'e}prez}}, \bibinfo {author} {\bibfnamefont {M.}~\bibnamefont {Meyer}}, \bibinfo {author} {\bibfnamefont {I.~N.}\ \bibnamefont {Meijer}}, \bibinfo {author} {\bibfnamefont {L.}~\bibnamefont {Tryputen}}, \bibinfo {author} {\bibfnamefont {S.}~\bibnamefont {Karwal}}, \bibinfo {author} {\bibfnamefont {M.}~\bibnamefont {Botifoll}},  \emph {et~al.},\ }\href {\doibase https://doi.org/10.1038/s41534-024-00826-9} {\bibfield  {journal} {\bibinfo  {journal} {npj Quantum Information}\ }\textbf {\bibinfo {volume} {10}},\ \bibinfo {pages} {32} (\bibinfo {year} {2024})}\BibitemShut {NoStop}%
\bibitem [{\citenamefont {Paquelet~Wuetz}\ \emph {et~al.}(2020)\citenamefont {Paquelet~Wuetz}, \citenamefont {Bavdaz}, \citenamefont {Yeoh}, \citenamefont {Schouten}, \citenamefont {Van Der~Does}, \citenamefont {Tiggelman}, \citenamefont {Sabbagh}, \citenamefont {Sammak}, \citenamefont {Almudever}, \citenamefont {Sebastiano} \emph {et~al.}}]{paquelet2020multiplexed}%
  \BibitemOpen
  \bibfield  {author} {\bibinfo {author} {\bibfnamefont {B.}~\bibnamefont {Paquelet~Wuetz}}, \bibinfo {author} {\bibfnamefont {P.}~\bibnamefont {Bavdaz}}, \bibinfo {author} {\bibfnamefont {L.}~\bibnamefont {Yeoh}}, \bibinfo {author} {\bibfnamefont {R.}~\bibnamefont {Schouten}}, \bibinfo {author} {\bibfnamefont {H.}~\bibnamefont {Van Der~Does}}, \bibinfo {author} {\bibfnamefont {M.}~\bibnamefont {Tiggelman}}, \bibinfo {author} {\bibfnamefont {D.}~\bibnamefont {Sabbagh}}, \bibinfo {author} {\bibfnamefont {A.}~\bibnamefont {Sammak}}, \bibinfo {author} {\bibfnamefont {C.~G.}\ \bibnamefont {Almudever}}, \bibinfo {author} {\bibfnamefont {F.}~\bibnamefont {Sebastiano}},  \emph {et~al.},\ }\href {\doibase https://doi.org/10.1038/s41534-020-0274-4} {\bibfield  {journal} {\bibinfo  {journal} {npj Quantum Information}\ }\textbf {\bibinfo {volume} {6}},\ \bibinfo {pages} {43} (\bibinfo {year} {2020})}\BibitemShut {NoStop}%
\bibitem [{\citenamefont {Myronov}\ \emph {et~al.}(2023)\citenamefont {Myronov}, \citenamefont {Kycia}, \citenamefont {Waldron}, \citenamefont {Jiang}, \citenamefont {Barrios}, \citenamefont {Bogan}, \citenamefont {Coleridge},\ and\ \citenamefont {Studenikin}}]{myronov2023holes}%
  \BibitemOpen
  \bibfield  {author} {\bibinfo {author} {\bibfnamefont {M.}~\bibnamefont {Myronov}}, \bibinfo {author} {\bibfnamefont {J.}~\bibnamefont {Kycia}}, \bibinfo {author} {\bibfnamefont {P.}~\bibnamefont {Waldron}}, \bibinfo {author} {\bibfnamefont {W.}~\bibnamefont {Jiang}}, \bibinfo {author} {\bibfnamefont {P.}~\bibnamefont {Barrios}}, \bibinfo {author} {\bibfnamefont {A.}~\bibnamefont {Bogan}}, \bibinfo {author} {\bibfnamefont {P.}~\bibnamefont {Coleridge}}, \ and\ \bibinfo {author} {\bibfnamefont {S.}~\bibnamefont {Studenikin}},\ }\href {\doibase https://doi.org/10.1002/smsc.202200094} {\bibfield  {journal} {\bibinfo  {journal} {Small Science}\ }\textbf {\bibinfo {volume} {3}},\ \bibinfo {pages} {2200094} (\bibinfo {year} {2023})}\BibitemShut {NoStop}%
\bibitem [{\citenamefont {Huang}\ and\ \citenamefont {Das~Sarma}(2024)}]{huang2024understanding}%
  \BibitemOpen
  \bibfield  {author} {\bibinfo {author} {\bibfnamefont {Y.}~\bibnamefont {Huang}}\ and\ \bibinfo {author} {\bibfnamefont {S.}~\bibnamefont {Das~Sarma}},\ }\href {\doibase https://doi.org/10.1103/PhysRevB.109.125405} {\bibfield  {journal} {\bibinfo  {journal} {Physical Review B}\ }\textbf {\bibinfo {volume} {109}},\ \bibinfo {pages} {125405} (\bibinfo {year} {2024})}\BibitemShut {NoStop}%
\bibitem [{\citenamefont {Last}\ and\ \citenamefont {Thouless}(1971)}]{last1971percolation}%
  \BibitemOpen
  \bibfield  {author} {\bibinfo {author} {\bibfnamefont {B.~J.}\ \bibnamefont {Last}}\ and\ \bibinfo {author} {\bibfnamefont {D.~J.}\ \bibnamefont {Thouless}},\ }\href {\doibase https://doi.org/10.1103/PhysRevLett.27.1719} {\bibfield  {journal} {\bibinfo  {journal} {Physical Review Letters}\ }\textbf {\bibinfo {volume} {27}},\ \bibinfo {pages} {1719} (\bibinfo {year} {1971})}\BibitemShut {NoStop}%
\bibitem [{\citenamefont {Shklovskiĭ}\ and\ \citenamefont {Éfros}(1975)}]{shklovskii1975percolation}%
  \BibitemOpen
  \bibfield  {author} {\bibinfo {author} {\bibfnamefont {B.~I.}\ \bibnamefont {Shklovskiĭ}}\ and\ \bibinfo {author} {\bibfnamefont {A.~L.}\ \bibnamefont {Éfros}},\ }\href {\doibase https://doi.org/10.1070/PU1975v018n11ABEH005233} {\bibfield  {journal} {\bibinfo  {journal} {Soviet Physics Uspekhi}\ }\textbf {\bibinfo {volume} {18}},\ \bibinfo {pages} {845} (\bibinfo {year} {1975})}\BibitemShut {NoStop}%
\bibitem [{\citenamefont {Tracy}\ \emph {et~al.}(2009)\citenamefont {Tracy}, \citenamefont {Hwang}, \citenamefont {Eng}, \citenamefont {Ten~Eyck}, \citenamefont {Nordberg}, \citenamefont {Childs}, \citenamefont {Carroll}, \citenamefont {Lilly},\ and\ \citenamefont {Das~Sarma}}]{tracy2009observation}%
  \BibitemOpen
  \bibfield  {author} {\bibinfo {author} {\bibfnamefont {L.}~\bibnamefont {Tracy}}, \bibinfo {author} {\bibfnamefont {E.}~\bibnamefont {Hwang}}, \bibinfo {author} {\bibfnamefont {K.}~\bibnamefont {Eng}}, \bibinfo {author} {\bibfnamefont {G.}~\bibnamefont {Ten~Eyck}}, \bibinfo {author} {\bibfnamefont {E.}~\bibnamefont {Nordberg}}, \bibinfo {author} {\bibfnamefont {K.}~\bibnamefont {Childs}}, \bibinfo {author} {\bibfnamefont {M.}~\bibnamefont {Carroll}}, \bibinfo {author} {\bibfnamefont {M.}~\bibnamefont {Lilly}}, \ and\ \bibinfo {author} {\bibfnamefont {S.}~\bibnamefont {Das~Sarma}},\ }\href {\doibase https://doi.org/10.1103/PhysRevB.79.235307} {\bibfield  {journal} {\bibinfo  {journal} {Physical Review B—Condensed Matter and Materials Physics}\ }\textbf {\bibinfo {volume} {79}},\ \bibinfo {pages} {235307} (\bibinfo {year} {2009})}\BibitemShut {NoStop}%
\bibitem [{\citenamefont {Fogelholm}(1980)}]{fogelholm1980conductivity}%
  \BibitemOpen
  \bibfield  {author} {\bibinfo {author} {\bibfnamefont {R.}~\bibnamefont {Fogelholm}},\ }\href {\doibase https://doi.org/10.1088/0022-3719/13/23/001} {\bibfield  {journal} {\bibinfo  {journal} {Journal of Physics C: Solid State Physics}\ }\textbf {\bibinfo {volume} {13}},\ \bibinfo {pages} {L571} (\bibinfo {year} {1980})}\BibitemShut {NoStop}%
\bibitem [{\citenamefont {Kim}\ \emph {et~al.}(2017)\citenamefont {Kim}, \citenamefont {Tyryshkin},\ and\ \citenamefont {Lyon}}]{kim2017annealing}%
  \BibitemOpen
  \bibfield  {author} {\bibinfo {author} {\bibfnamefont {J.-S.}\ \bibnamefont {Kim}}, \bibinfo {author} {\bibfnamefont {A.~M.}\ \bibnamefont {Tyryshkin}}, \ and\ \bibinfo {author} {\bibfnamefont {S.~A.}\ \bibnamefont {Lyon}},\ }\href {\doibase https://doi.org/10.1063/1.4979035} {\bibfield  {journal} {\bibinfo  {journal} {Applied Physics Letters}\ }\textbf {\bibinfo {volume} {110}} (\bibinfo {year} {2017}),\ https://doi.org/10.1063/1.4979035}\BibitemShut {NoStop}%
\bibitem [{\citenamefont {Lodari}\ \emph {et~al.}(2021)\citenamefont {Lodari}, \citenamefont {Hendrickx}, \citenamefont {Lawrie}, \citenamefont {Hsiao}, \citenamefont {Vandersypen}, \citenamefont {Sammak}, \citenamefont {Veldhorst},\ and\ \citenamefont {Scappucci}}]{lodari2021low}%
  \BibitemOpen
  \bibfield  {author} {\bibinfo {author} {\bibfnamefont {M.}~\bibnamefont {Lodari}}, \bibinfo {author} {\bibfnamefont {N.~W.}\ \bibnamefont {Hendrickx}}, \bibinfo {author} {\bibfnamefont {W.~I.}\ \bibnamefont {Lawrie}}, \bibinfo {author} {\bibfnamefont {T.-K.}\ \bibnamefont {Hsiao}}, \bibinfo {author} {\bibfnamefont {L.~M.}\ \bibnamefont {Vandersypen}}, \bibinfo {author} {\bibfnamefont {A.}~\bibnamefont {Sammak}}, \bibinfo {author} {\bibfnamefont {M.}~\bibnamefont {Veldhorst}}, \ and\ \bibinfo {author} {\bibfnamefont {G.}~\bibnamefont {Scappucci}},\ }\href {\doibase https://doi.org/10.1088/2633-4356/abcd82} {\bibfield  {journal} {\bibinfo  {journal} {Materials for Quantum Technology}\ }\textbf {\bibinfo {volume} {1}},\ \bibinfo {pages} {011002} (\bibinfo {year} {2021})}\BibitemShut {NoStop}%
\bibitem [{\citenamefont {Sabbagh}\ \emph {et~al.}(2019)\citenamefont {Sabbagh}, \citenamefont {Thomas}, \citenamefont {Torres}, \citenamefont {Pillarisetty}, \citenamefont {Amin}, \citenamefont {George}, \citenamefont {Singh}, \citenamefont {Budrevich}, \citenamefont {Robinson}, \citenamefont {Merrill} \emph {et~al.}}]{sabbagh2019quantum}%
  \BibitemOpen
  \bibfield  {author} {\bibinfo {author} {\bibfnamefont {D.}~\bibnamefont {Sabbagh}}, \bibinfo {author} {\bibfnamefont {N.}~\bibnamefont {Thomas}}, \bibinfo {author} {\bibfnamefont {J.}~\bibnamefont {Torres}}, \bibinfo {author} {\bibfnamefont {R.}~\bibnamefont {Pillarisetty}}, \bibinfo {author} {\bibfnamefont {P.}~\bibnamefont {Amin}}, \bibinfo {author} {\bibfnamefont {H.}~\bibnamefont {George}}, \bibinfo {author} {\bibfnamefont {K.}~\bibnamefont {Singh}}, \bibinfo {author} {\bibfnamefont {A.}~\bibnamefont {Budrevich}}, \bibinfo {author} {\bibfnamefont {M.}~\bibnamefont {Robinson}}, \bibinfo {author} {\bibfnamefont {D.}~\bibnamefont {Merrill}},  \emph {et~al.},\ }\href {\doibase https://doi.org/10.1103/PhysRevApplied.12.014013} {\bibfield  {journal} {\bibinfo  {journal} {Physical Review Applied}\ }\textbf {\bibinfo {volume} {12}},\ \bibinfo {pages} {014013} (\bibinfo {year} {2019})}\BibitemShut {NoStop}%
\bibitem [{\citenamefont {Paysen}\ \emph {et~al.}(2024)\citenamefont {Paysen}, \citenamefont {Capellini}, \citenamefont {Talamas~Simola}, \citenamefont {Di~Gaspare}, \citenamefont {De~Seta}, \citenamefont {Virgilio},\ and\ \citenamefont {Trampert}}]{paysen2024three-dimensional}%
  \BibitemOpen
  \bibfield  {author} {\bibinfo {author} {\bibfnamefont {E.}~\bibnamefont {Paysen}}, \bibinfo {author} {\bibfnamefont {G.}~\bibnamefont {Capellini}}, \bibinfo {author} {\bibfnamefont {E.}~\bibnamefont {Talamas~Simola}}, \bibinfo {author} {\bibfnamefont {L.}~\bibnamefont {Di~Gaspare}}, \bibinfo {author} {\bibfnamefont {M.}~\bibnamefont {De~Seta}}, \bibinfo {author} {\bibfnamefont {M.}~\bibnamefont {Virgilio}}, \ and\ \bibinfo {author} {\bibfnamefont {A.}~\bibnamefont {Trampert}},\ }\href {\doibase https://doi.org/10.1021/acsami.3c15546} {\bibfield  {journal} {\bibinfo  {journal} {ACS Applied Materials \& Interfaces}\ }\textbf {\bibinfo {volume} {16}},\ \bibinfo {pages} {4189} (\bibinfo {year} {2024})}\BibitemShut {NoStop}%
\bibitem [{\citenamefont {Das~Sarma}\ and\ \citenamefont {Hwang}(2014)}]{dassarma2014mobility}%
  \BibitemOpen
  \bibfield  {author} {\bibinfo {author} {\bibfnamefont {S.}~\bibnamefont {Das~Sarma}}\ and\ \bibinfo {author} {\bibfnamefont {E.~H.}\ \bibnamefont {Hwang}},\ }\href {\doibase https://doi.org/10.1103/PhysRevB.90.035425} {\bibfield  {journal} {\bibinfo  {journal} {Physical Review B}\ }\textbf {\bibinfo {volume} {90}},\ \bibinfo {pages} {035425} (\bibinfo {year} {2014})}\BibitemShut {NoStop}%
\bibitem [{\citenamefont {Costa}\ and\ \citenamefont {Scappucci}(2024)}]{dataset}%
  \BibitemOpen
  \bibfield  {author} {\bibinfo {author} {\bibfnamefont {D.}~\bibnamefont {Costa}}\ and\ \bibinfo {author} {\bibfnamefont {G.}~\bibnamefont {Scappucci}},\ }\href {\doibase https://doi.org/10.4121/a4fce2ed-262f-4db9-866b-661c0e002671} {\bibfield  {journal} {\bibinfo  {journal} {4TU.ResearchData. Dataset}\ } (\bibinfo {year} {2024}),\ https://doi.org/10.4121/a4fce2ed-262f-4db9-866b-661c0e002671}\BibitemShut {NoStop}%
\end{thebibliography}
\end{document}


\title{Supplementary material: Reducing disorder in Ge quantum wells by using thick SiGe barriers}

\author{Davide Costa}
\affiliation{QuTech and Kavli Institute of Nanoscience, Delft University of Technology, Lorentzweg 1, 2628 CJ Delft, Netherlands}
\author{Lucas E. A. Stehouwer}
\affiliation{QuTech and Kavli Institute of Nanoscience, Delft University of Technology, Lorentzweg 1, 2628 CJ Delft, Netherlands}
\author{Yi Huang}
\affiliation{Condensed Matter Theory Center and Joint Quantum Institute, Department of Physics, University of Maryland, College Park, Maryland 20742, USA}
\author{Sara Martí-Sánchez}
\affiliation{Catalan Institute of Nanoscience and Nanotechnology (ICN2), CSIC and BIST, Campus UAB, Bellaterra, 08193 Barcelona, Catalonia, Spain}
\author{Davide Degli Esposti}
\affiliation{QuTech and Kavli Institute of Nanoscience, Delft University of Technology, Lorentzweg 1, 2628 CJ Delft, Netherlands}
\author{Jordi Arbiol}
\affiliation{Catalan Institute of Nanoscience and Nanotechnology (ICN2), CSIC and BIST, Campus UAB, Bellaterra, 08193 Barcelona, Catalonia, Spain}
\affiliation{ICREA, Pg. Lluís Companys 23, 08010 Barcelona, Catalonia, Spain}
\author{Giordano Scappucci}
\email{g.scappucci@tudelft.nl}
\affiliation{QuTech and Kavli Institute of Nanoscience, Delft University of Technology, Lorentzweg 1, 2628 CJ Delft, Netherlands}

\date{\today}
\pacs{}

\maketitle

\section{Scanning transmission electron microscopy}

We evaluate the thickness of the quantum well $w$ and the sharpness of the top and bottom interfaces by fitting the intensity profile of the high angle annular dark field (HAADF) scanning transmission electron microscopy (STEM) image in Fig.~1(b) of the main text  with a sigmoid function:
\begin{equation}
    I(x) = \frac{1}{1 + e^{\frac{x_{top}-x}{\tau_{top}}}} + \frac{1}{1 + e^{\frac{x - x_{bottom}}{\tau_{bottom}}}},
\end{equation}

In this function, $x_{top}$ and $x_{bottom}$ are the inflection points of the sigmoid, defining the positions of the top (Ge$\rightarrow$SiGe) and bottom (SiGe$\rightarrow$Ge) interfaces of the quantum well, and $\tau_{top}$ and $\tau_{bottom}$ are the their characteristic lengths.  The thickness of the quantum well $w$ is given by the distance between $x_{top}$ and $x_{bottom}$, the parameter $4\tau$ corresponds to the length over which the intensity profile changes from 0.12 to 0.88 of the asymptotic value. 

\begin{figure}[H]
    \centering
	\includegraphics[width=70mm]{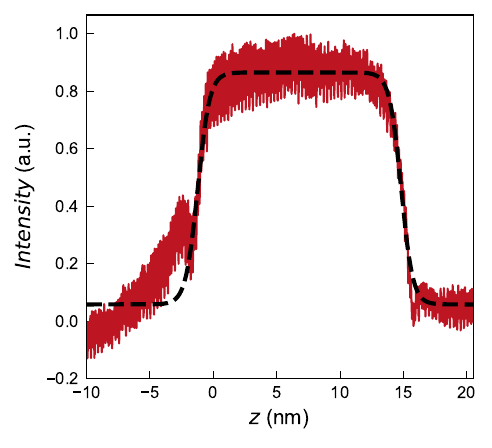}
	\caption{Averaged intensity profile of the HAADF-STEM image of the heterostructure active layers (red curve) with a Sigmoid fit to extract characteristic dimensions of the QW (dashed black curve).}  
\label{fig:S1}
\end{figure}

For the intensity profile and fit reported in Fig.~\ref{fig:S1}, we find a quantum well thickness $w$ of $(16.04 \pm 0.03) ~\mathrm{nm}$ and a top and bottom interface sharpness $4\tau$ of $(1.88 \pm 0.06) ~\mathrm{nm}$ and $(1.83 \pm 0.05) ~\mathrm{nm}$, respectively.  The quantum well thickness is similar to previously grown Ge/SiGe heterostructures on Si or Ge wafers \cite{lodari2021low,stehouwer2023germanium}. The intensity profile of the top and bottom interfaces shows Si accumulation (or Ge depletion) that makes the intensity profile deviate from a simple sigmoid model. Further insights may be gained by advanced electron microscopy \cite{paysen2024three-dimensional} or atom probe tomography \cite{paquelet2022atomic} characterization.

\section{Scattering theory and fit of mobility-density curves}

The hole mobility of Ge/SiGe quantum wells is calculated using Boltzmann transport theory at $T=0$. The relevant material parameters are provided below. The in-plane effective hole mass in Ge is $m=0.055 m_{\mathrm{0}}$ where $m_{\mathrm{0}}$ is the free electron mass, while the out-of-plane effective hole mass in Ge is $m=0.22 m_{\mathrm{0}}$ \cite{lodari2019light,terrazos2021theory}, specifically applicable for the scenario of a compressively strained Ge quantum well embedded in strain-relaxed Si$_{0.17}$Ge$_{0.83}$. The overall quantum degeneracy is $g=2$ (comprising a spin degeneracy of $g_{\mathrm{s}} = 2$ and a valley degeneracy of $g_{\mathrm{v}} = 1$, given the large splitting between heavy and light hole bands \cite{sammak2019shallow}). The dielectric constant for germanium is $\epsilon=16$ and the thickness of the quantum well is $w=16 ~\mathrm{nm}$. The thickness of the top SiGe barrier is $d=135 ~\mathrm{nm}$. The thickness of the Al$_2$O$_3$ oxide layer on top of the SiGe barrier is $d_{\mathrm{o}}=30 ~\mathrm{nm}$. The valence band offset (representing the potential barrier height) between the strained Ge and the strain-relaxed SiGe barrier is $V_{\mathrm{0}} = 150 ~\mathrm{meV}$ \cite{sammak2019shallow}. The lattice constant of Si (Ge) is $5.4$ {\AA} ($5.7$ {\AA}). The thickness of the lower SiGe barrier is $L_{\mathrm{buf}}=1.5 ~\mathrm{\upmu m}$, i.e. much greater than the the top barrier and effectively irrelevant for this discussion. \\
In the calculation of mobility, we use the wavefunction of a finite potential well and take into account both the gate screening effect and the local field correction \cite{huang2024understanding}. The mobility is evaluated using the Born approximation
\begin{equation}
   \frac{1}{\tau} =\frac{4m}{\pi \hbar^3} \int \limits_0^{2k}\frac{dq}{\sqrt{4k^2 - q^2}} \qty(\frac{q}{2k})^2 \ev{\abs{U(q)}^2}\,,
\label{eq:1_tau}
\end{equation}
where $U(q)$ is the screened potential of a given scattering source.
The scattering rate contributed by uniform background charged impurities reads
\begin{align}
    \frac{1}{\tau_{\mathrm{BI}}} = \frac{N_{\mathrm{b}}}{k_\mathrm{F}} \frac{2\pi\hbar}{m} \qty(\frac{2}{g})^2 f_b(s),
\end{align}
where $N_{\mathrm{b}}$ is the background impurity concentration and $k_\mathrm{F} = \sqrt{4\pi p/g_{\mathrm{s}}g_{\mathrm{v}}}$ is the Fermi wavelength.
$f_{\mathrm{b}}(s)$ is a dimensionless function given by
\begin{align}
    f_{\mathrm{b}}(s) &= \int_0^1 \frac{x dx}{\sqrt{1-x^2} (x/s + 1)^2} \\
    &= \begin{cases}
        \frac{s^2}{s^2 - 1} - \frac{s^2 \mathrm{sec}^{-1}(s)}{(s^2 - 1)^{3/2}}, \, &s\geq 1, \\
        \frac{s^2}{s^2 - 1} + \frac{s^2 \ln(\sqrt{s^{-2} - 1} + s^{-1})}{(1 - s^2)^{3/2}}, \, &s<1.
    \end{cases}
\end{align}
where $s$ is the screening parameter, equal to $q_{\mathrm{TF}}/2k_{\mathrm{F}}$, with 
\begin{align}
    q_{\mathrm{TF}} = \qty[g_{\mathrm{s}}g_{\mathrm{v}}(m/m_{\mathrm{0}})/\epsilon] \qty(e^2m_{\mathrm{0}}/4\pi \epsilon_{\mathrm{0}}\hbar^2).
\end{align}
For Ge/SiGe quantum wells, the numerical value of the screening parameter is $s^2 \approx 0.67/p$, where $p$ is in units of $10^{11} ~\mathrm{cm^{-2}}$. This means that the screening is weaker in Ge/SiGe hole systems, compared with $s^2 \approx 116/n$ in Si/SiGe electron systems and $s^2 \approx 1.5/n$ in GaAs/AlGaAs electron systems. In the strong screening limit $s \gg 1$, we have $f_{\mathrm{b}}(s) \approx 1$; while in the weak screening limit $s \ll 1$, we have $f_{\mathrm{b}}(s) \approx -s^2 \ln s$.

Using an infinite potential well wavefunction, we then calculate the interface roughness (IR) scattering rate
\begin{align}
    \frac{1}{\tau_{\mathrm{IR}}} = \frac{ 2 \pi^4 \hbar}{m}\qty(\frac{m}{m_{\mathrm{z}}})^2 \frac{\Delta^2 \Lambda^2}{w^6} f_{\mathrm{i}}(s,a_{\mathrm{IR}}),
\end{align}
where $\Delta$ is the typical height and $\Lambda$ is the lateral size of the interface roughness.
The dimensionless function is given by
\begin{align}
    f_{\mathrm{i}}(s,a_{\mathrm{IR}}) = \int_{0}^{1} \frac{2x^4 e^{-x^2 a_{\mathrm{IR}}^2}\,dx}{\sqrt{1-x^2}(x+s)^2}, 
\end{align}
where $a_{\mathrm{IR}}=k_F \Lambda$. In the strong screening limit $s \gg 1$ we have 
\begin{align}
    f_{\mathrm{i}}(s \gg 1,a_{\mathrm{IR}}) \approx 
    \begin{cases}
        \frac{3\pi}{8 s^2}, \, &a_{\mathrm{IR}} \ll 1,\\
        \frac{3\sqrt{\pi}}{4 s^2 a_{\mathrm{IR}}^5} \, &a_{\mathrm{IR}} \gg 1.
    \end{cases}
\end{align}
In the weak screening limit $s \ll 1$ we have
\begin{align}
    f_{i}(s \ll 1,a_{\mathrm{IR}}) \approx 
    \begin{cases}
        \frac{\pi}{2}, \, &a_{\mathrm{IR}} \ll 1,\\
        \frac{\sqrt{\pi}}{2 a_{\mathrm{IR}}^3}, \, &a \gg 1, \, sa_{\mathrm{IR}} \ll 1.
    \end{cases}
\end{align}
The finite-potential-well wave function modifies the above expression by a factor of $\tilde{E}^2$ given by
\begin{align}
    \tilde E \approx 1 - \frac{4}{\pi \tilde V^{1/2}} + \frac{12}{\pi^2 \tilde V} + \mathcal{O}\qty(\tilde V^{-\tfrac{3}{2}}),
\end{align}
where $\tilde{E} = E/E_0$ and $\tilde V = (V_0-E_{\mathrm{F}})/E_0$. 
$E$ and $E_0$ are the finite- and infinite-potential-well ground state energies, respectively. $V_0$ is the potential barrier. In our case of a $16 ~\mathrm{nm}$ Ge quantum well around a hole density $p = 1\times10^{11} ~\mathrm{cm^{-2}}$, we have $\tilde V \approx 22$, which corresponds to $\tilde E \approx 0.77$. The relevant parameters extracted from the mobility fit through the BI (dashed black line) and IR (dotted black line) models are $N_{\mathrm{b}} = 3.8\times10^{13} ~\mathrm{cm^{-3}}$, $\Delta = 10$ {\AA}, and $\Lambda = 28$ {\AA}.

To further inspect the scattering mechanisms, we also explore the scattering rate contributed by remote charged impurities, which reads
\begin{align}
    \frac{1}{\tau_{\mathrm{RI}}} = n_{\mathrm{r}} \frac{2\pi\hbar}{m} \qty(\frac{2}{g})^2 f_{\mathrm{r}}(s, a_{\mathrm{RI}}),
\end{align}
where $n_{\mathrm{r}}$ is the remote impurities concentration. $f_{\mathrm{r}}(s,a_{\mathrm{RI}})$ is a dimensionless function given by
\begin{align}\label{eq:f0_no_gate}
    f_{\mathrm{r}}(s,a_{\mathrm{RI}}) = \int_{0}^{1} \frac{2x^2 e^{-2a_{\mathrm{RI}}x}\,dx}{\sqrt{1-x^2}(x/s+1)^2},
\end{align}
where $a_{\mathrm{RI}} = 2k_{\mathrm{F}} d$.
For far away impurities $a_{\mathrm{RI}} \gg 1$ such that $sa_{\mathrm{RI}} \gg 1$, we have $f_{\mathrm{r}}(s,a_{\mathrm{RI}}) \approx (2a_{\mathrm{RI}}^3)^{-1}$. In the regime where $a_{\mathrm{RI}} \gg 1$ and under the condition of weak screening with $s \ll 1$ and $sa_{\mathrm{RI}} \ll 1$, the function $f_{\mathrm{r}}(s,a_{\mathrm{RI}})$ can be approximated as $s^2/a_{\mathrm{RI}}$. For our case $d_{\mathrm{o}} \ll d$ such that remote impurities are relatively close to the gate, and $sa_{\mathrm{RI}}=q_{\mathrm{TF}} d = 17.6 \gg 1$ for $d=135 ~\mathrm{nm}$ ($q_{\mathrm{TF}} d = 7.2 \gg 1$ for $d=55 ~\mathrm{nm}$), so we have the following expressions for the scattering rate with gate screening \cite{huang2024understanding}
\begin{gather}\label{eq:rate1_gate}
    \frac{1}{\tau_{\mathrm{RI}}} =  \frac{\pi\hbar n_r }{8m k_{\mathrm{F}}^3 d^3 } \qty(1 + \frac{d^3}{(2d_{\mathrm{g}} - d)^3} - \frac{2d^3}{d_{\mathrm{g}}^3}) \qty(\frac{2}{g})^2 , 
\end{gather}
where $d_{\mathrm{g}} = d+d_{\mathrm{o}}+w/2$ is the distance from the gate to the center of the quantum well. The mobility $\mu_{\mathrm{RI}}$ has therefore a quasi-cubic dependence with the thickness of the top SiGe barrier $d$. This strong dependence allows to evaluate the relative contribution of remote impurities to scattering compared to the other mechanism. If remote impurities were the primary scattering mechanism, increasing the barrier thickness from $55 ~\mathrm{nm}$ \cite{stehouwer2023germanium} to $135 ~\mathrm{nm}$ would have resulted in a more than eightfold increase in mobility. Instead, the mobility increased by a factor of $1.6$ in the heterostructure with the 135 nm thick SiGe barrier, pointing to a marginal role of remote impurities as scattering limiting mechanism.

\section{Conductivity-density curves}
\begin{figure}[H]
    \centering
	\includegraphics[width=85mm]{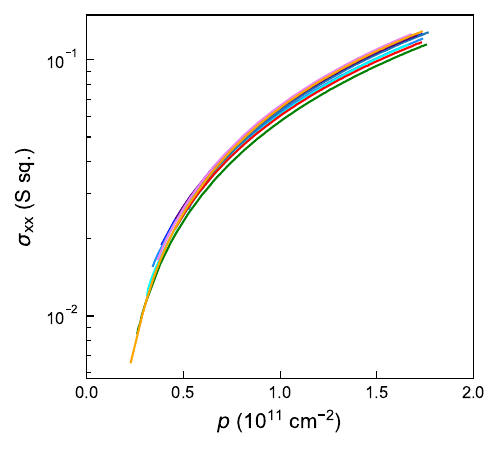}
	\caption{Longitudinal conductivity $\sigma_{\mathrm{xx}}$ as a function of Hall density $p$ for the nine investigated heterostructure field effect transistors (colored lines).}  
\label{fig:S2}
\end{figure}

In Fig.~\ref{fig:S2} we show the conductivity density curves for all nine heterostructure field effect transistors. The orange curve, measured to the lowest density, was used for the percolation density analysis in Fig.~3 of the main text. 

